\documentclass{PoS}

\title{Analysis of the $N_f=2+1$ lattice QCD results on the lowest-lying baryon masses using covariant ChPT}

\ShortTitle{Analysis of the $N_f=2+1$ lattice QCD results on the lowest-lying baryon masses using covariant ChPT}

\author{\speaker{Jorge MARTIN CAMALICH}\\
Departamento de F\'{\i}sica Te\'orica and IFIC, Universidad de
Valencia-CSIC, Spain \\
        E-mail: \email{camalich@ific.uv.es}}

\author{Lisheng GENG\\
       School of Physics and Nuclear Energy Engineering, Beihang University, Beijing 100191, China\\
       Physik Department, Technische Universit\" at M\"unchen, D-85747 Garching, Germany\\
       E-mail: \email{lisheng.geng@ph.tum.de}}

\author{Manuel Jos\'e VICENTE VACAS\\
      Departamento de F\'{\i}sica Te\'orica and IFIC, Universidad de
      Valencia-CSIC, Spain  \\
       E-mail: \email{vicente@ific.uv.es}}

\abstract{We review recent progress in the understanding of low-energy baryon structure by means
of chiral perturbation theory. In particular, we discuss the application of this formalism to the
description of the quark mass dependence of recent Lattice QCD results on the masses. We present
the chiral extrapolation of those of the PACS-CS and LHP collaborations and we predict the baryonic
sigma-terms.}

\FullConference{The XXVIII International Symposium on Lattice Filed Theory\\
                 June 14-19,2010\\
                 Villasimius, Sardinia Italy}

\begin{document}

Baryons are physical objects of great interest. Their properties and interactions are essential to understand those of the atomic nuclei or of more exotic kinds of systems
like the strange matter, which is believed to play a role in the macroscopic properties of astrophysical objects, e.g. neutron stars. On the other hand, baryon phenomenology allows to study the non-perturbative regime of Quantum Chromodynamics (QCD). It is a great scientific endeavor to understand the extremely rich spectroscopy and structure of baryons directly from the few parameters of QCD, namely the strong coupling constant and quark masses. Additionally, their weak decays and reactions provide information on the flavor structure of the electroweak interactions that eventually may point out departures from the Standard Model (SM) predictions in baryonic observables.

Experiments on baryon spectroscopy, structure, decays or reactions are currently taking data or are planned in laboratories like CERN-SPS, GSI, etc, and new facilities will be soon available at J-PARC, TJNAF, LNF, etc. Moreover, the last few years have witnessed an impressive development in the Lattice QCD (LQCD) description of several observables and realistic results on baryon structure are starting to appear~\cite{Jansen:2008vs} . On the other hand the investigation of baryon phenomenology by means of the low-energy effective field theory of QCD, namely chiral perturbation theory ($\chi$PT)~\cite{Gasser:1984gg,Gasser:1987rb} , has been troubled for a very long time because of several conceptual and technical problems like
the poor convergence of the perturbative (chiral) series in teh heavy-baryon (HB) approach~\cite{Jenkins:1990jv} or the effects of the lowest-lying decuplet resonances. Recently, we have shown that a fairly good convergence is possible using a Lorentz covariant approach with a consistent power counting~\cite{Fuchs:2003qc} that systematically incorporates the decuplet resonances~\cite{Pascalutsa:2000kd,Geng:2009hh}. A model-independent understanding of diferent properties including the magnetic moments of the baryon-octet~\cite{Geng:2008mf,Geng:2009hh}, the electromagnetic structure
of the decuplet resonances~\cite{Geng:2009ys} and the hyperon vector coupling $f_1(0)$~\cite{Geng:2009ik}, has been successfully achieved. Finally, it is worth noticing that this approach has been extended to heavy-light systems~\cite{Geng:2010vw}. In this work we review an application that has been worked out in connection with lattice simulations, namely the extrapolation of LQCD results on the baryon masses~\cite{MartinCamalich:2010fp}.

\begin{figure}[t]
\centering
\includegraphics[width=12cm]{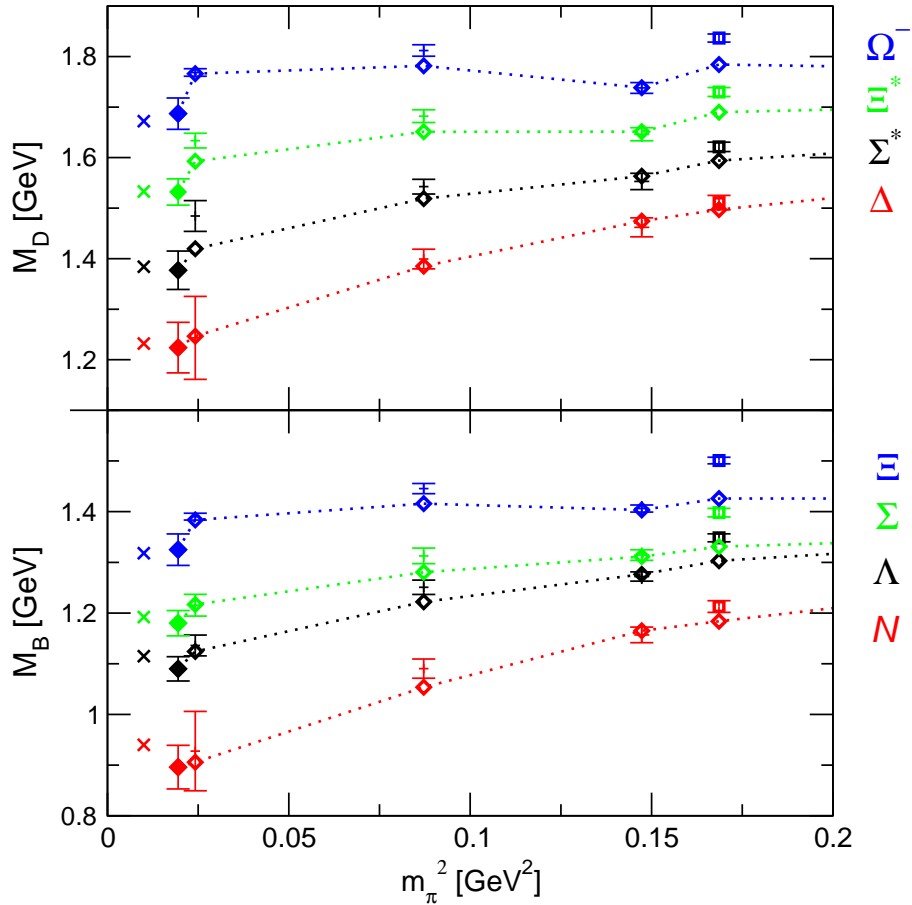}
\caption{Extrapolation of the PACS-CS results~\cite{Aoki:2008sm} on the lowest-lying baryon masses within the 
covariant formulation of $SU(3)_F$-B$\chi$PT up to NLO. The LQCD points used in the 
fit are represented with the corresponding error bars which do not include the correlated uncertainties. The lattice points in $m_\pi^2\simeq0.15$ GeV$^2$ involve a lighter strange quark mass. The diamonds denote our  
results after the fit and they are connected by a dotted line added to guide the eye.
The boxes are lattice points not included in the fit (heavier kaon mass) and the filled diamonds are the extrapolated values which are to be compared with experimental data (crosses). The latter are slightly shifted for a better comparison with the extrapolation results. \label{fig_graph}}
\end{figure}

In the last decades there has been a sustained interest in the description of the lowest-lying baryon mass spectrum by means of $SU(3)_F$-B$\chi$PT (see e.g. Refs.~\cite{Jenkins:1991ts,Bernard:1993nj,Borasoy:1996bx,WalkerLoud:2004hf,Tiburzi:2004rh,Lehnhart:2004vi,MartinCamalich:2010fp}). The chiral corrections to the Gell-Mann-Okubo baryon octet mass relation and Gell-Mann's decuplet equal spacing rules (we denote generically as GMO),
\begin{eqnarray}
&&\hspace{0.9cm}3M_\Lambda+M_\Sigma-2M_N-2M_\Xi=0\label{Eq:GMOBar}\\
&&M_{\Sigma^*}-M_\Delta=M_{\Xi^*}-M_{\Sigma^*}=M_{\Omega^-}-M_{\Xi^*}\label{Eq:DES} 
\end{eqnarray}
have received special attention. More specifically, the GMO relations, which are recovered in $\chi$PT at LO, are known to work with an accuracy of $\sim7$ MeV. A puzzling and not yet well understood feature of the leading chiral corrections is that they preserve the GMO equations within $\sim10$ MeV whereas the corrections to any of the individual baryon masses are of order $\sim$100-1000 MeV~\cite{Jenkins:1991ts}. Another interesting aspect is that the analysis of the baryon masses provides hints on their scalar structure, i.e. the sigma terms~\cite{Pagels:1974se,Bernard:1993nj}. These magnitudes, besides providing a measure of the explicit symmetry breaking and of the meson-cloud contribution to the baryon masses~\cite{Pagels:1974se}, are relevant for other areas of phenomenology~\cite{Ellis:2008hf,Giedt:2009mr}. 

\begin{table*}[t]
\renewcommand{\arraystretch}{1.3}     
\setlength{\tabcolsep}{0.1cm}
\centering
\caption{Values of the LECs in the baryon-octet sector from the fits to the experimental, the PACS-CS~\cite{Aoki:2008sm} and the LHP~\cite{WalkerLoud:2008bp} results on the baryon masses using Lorentz covariant B$\chi$PT up to NLO. \label{Table:LECsOct}}
\begin{tabular}{c|ccc|cc|}
\cline{2-6}
&$M_{B0}$ [GeV]&$b_0$ [GeV$^{-1}$]&$M_{B0}^{eff}$ [GeV]&$b_D$ [GeV$^{-1}$]&$b_F$ [GeV$^{-1}$]\\
\hline
\multicolumn{1}{|c|}{Expt.}&-&-&1.840(5)&0.199(4)&$-$0.530(2)\\
\multicolumn{1}{|c|}{PACS-CS}&0.756(32)&$-0.978(38)$&1.76(7)&0.190(24)&$-$0.519(19)\\
\multicolumn{1}{|c|}{LHP}&0.780(31)&$-1.044(45)$&1.85(8)&0.236(24)&$-$0.523(21)\\
\hline
\end{tabular}
\end{table*}

\begin{table*}[t]
\renewcommand{\arraystretch}{1.3}     
\setlength{\tabcolsep}{0.1cm}
\centering
\caption{Values of the LECs in the baryon-decuplet sector from the fits to the experimental, the PACS-CS~\cite{Aoki:2008sm} and the LHP~\cite{WalkerLoud:2008bp} results on the baryon masses using Lorentz covariant B$\chi$PT up to NLO. \label{Table:LECsDec}}
\begin{tabular}{c|ccc|c|}
\cline{2-5}
&$M_{T0}$ [GeV]&$t_0$ [GeV$^{-1}$]&$M_{T0}^{eff}$ [GeV]&$t_D$ [GeV$^{-1}$]\\
\hline
\multicolumn{1}{|c|}{Expt.}&-&-&1.519(2)&$-$0.694(2)\\
\multicolumn{1}{|c|}{PACS-CS}&954(37)&$-1.05(8)$&1.49(8)&$-$0.682(20)\\
\multicolumn{1}{|c|}{LHP}&944(42)&$-1.28(8)$&1.60(8)&$-$0.609(14)\\
\hline
\end{tabular}
\end{table*}

On the other hand, LQCD calculations of the lowest-lying baryon mass spectrum have been undertaken by different collaborations using $N_f=2+1$ dynamical actions with light quark masses close to the physical point~\cite{WalkerLoud:2008bp,Aoki:2008sm,Lin:2008pr,Durr:2008zz,Bietenholz:2010jr}. The LHP~\cite{WalkerLoud:2008bp} and PACS-CS~\cite{Aoki:2008sm} collaborations have reported tremendous difficulties to understand the quark mass dependence and the chiral extrapolation of their results within HB$\chi$PT. This problem has been recently revisited in Ref.~\cite{MartinCamalich:2010fp}. In sharp contrast with the results obtained using the heavy-baryon expansion, it has been found that a good description of the LQCD results can be achieved within the Lorentz covariant approach to $SU(3)_F$-B$\chi$PT up to NLO. Moreover, the values of the masses extrapolated to the physical point of quark masses are manifestly better at NLO than those obtained using the linear extrapolation given by the GMO approach at LO. The study of the results of the LHP collaboration~\cite{WalkerLoud:2008bp} confirm all these conclusions. 

In Fig.~\ref{fig_graph}, we show the quark mass dependence and extrapolation of the lowest-lying baryon masses in Lorentz covariant $\chi$PT for the case of the analysis of the PACS-CS results~\cite{Aoki:2008sm}. The improvement obtained at NLO in covariant $SU(3)_F$-B$\chi$PT, highlights the effect of the leading chiral non-analytical terms in the extrapolation even from light quark masses as small as those used by PACS-CS.~\cite{Aoki:2008sm} ($m_\pi\simeq156$ MeV). On the other hand, the comparison between the results obtained in covariant and HB results~\cite{MartinCamalich:2010fp} illustrates the importance of the relativistic corrections in the understanding of the dependence shown by the lattice simulations on the baryon masses at relatively heavy quark-masses. This is specially true for the extrapolation of the LHP results, that are quite far away from the physical point of quark masses ($m_\pi\gtrsim293$ MeV). 

An important issue concerns the determination of the LECs of $SU(3)_F$-B$\chi$PT using 2+1-flavor 
simulations. In Tables~\ref{Table:LECsOct} and~\ref{Table:LECsDec}, we compare the values of the LECs determined studying the experimental values of the baryon masses with those obtained when fitting the corresponding quark-mass dependence of the PACS-CS or the LHP results. Since the experimental data do not disentangle $M_{B0}$ ($M_{T0}$) from $b_0$ ($t_0$) for the baryon-octet (-decuplet), in the comparison with the experimental determinations we must consider the effective masses $M_{B0}^{eff}$ ($M_{T0}^{eff}$) instead of these LECs~\cite{MartinCamalich:2010fp}.

In the case of the baryon-octet masses, the values of the LECs determined using either of the two LQCD sets of results partially agree with each other and they both are consistent with those resulting from the experimental determination. This suggests a non-trivial consistency in the baryon-octet sector between the lattice actions employed by the two collaborations (at different lattice spacings) and the experimental information on the masses through covariant B$\chi$PT up to NLO of accuracy. For the masses of the decuplet-baryons, while the values obtained using the PACS-CS results agree with those determined with the experimental data, the fit to the LHP results presents a value of $t_D$ that is not consistent with the experimental one. Some problems on these LQCD results for the decuplet masses were already noticed by the LHP collaboration~\cite{WalkerLoud:2008bp}.

\begin{table}[t]
\renewcommand{\arraystretch}{1.3}     
\setlength{\tabcolsep}{0.3cm}
\centering
\caption{Predictions on the $\sigma_{\pi N}$ and $\sigma_{sN}$ terms (in MeV) of the baryon-octet in covariant $SU(3)_F$-B$\chi$PT by fitting the LECs to the PACS-CS~\cite{Aoki:2008sm} or LHP~\cite{WalkerLoud:2008bp} results.   \label{Table:ResSigmasB}}
\begin{tabular}{c|cc|}
\cline{2-3}
&PACS-CS&LHP\\
\hline
\multicolumn{1}{|c|}{$\sigma_{\pi N}$}&59(2)(17)&61(2)(21)\\
\multicolumn{1}{|c|}{$\sigma_{sN}$}&$-7$(23)(25)&$-4$(20)(25)\\
\hline
\end{tabular}
\end{table}

A reliable combination of LQCD and $\chi$PT becomes a powerful framework to understand hadron phenomenology from first principles and may have sound applications. An example in the scalar sector is given by the determination of the sigma terms from the analysis of the masses through the Hellman-Feynman theorem,
\begin{equation}
\sigma_{\pi \mathcal{B}}=m\frac{\partial M_\mathcal{B}}{\partial m}, \hspace{1cm}\sigma_{s \mathcal{B}}=m_s\frac{\partial M_\mathcal{B}}{\partial m_s}. \label{Eq:sigmasBH-F}
\end{equation}
In Table ~\ref{Table:ResSigmasB} we present the results on the $\sigma_{\pi N}$ and $\sigma_{sN}$ obtained after fitting the LECs to the PACS-CS and LHP results and with the uncertainties determined as it has been discussed above for the masses. It is interesting to note that the results of the analysis of the two collaborations are in agreement with each other. Moreover they both also agree with those of Ref.~\cite{Young:2009zb} which are obtained within the cut-off renormalized B$\chi$PT. 

\section{Conclusions}

In summary, we have explored the applicability of the covariant formulation of $SU(3)_F$-B$\chi$PT within the EOMS scheme to analyze current 2+1-flavor
LQCD data, i.e. the results of the PACS-CS and LHP collaborations on the baryon masses. In contrast with the problems found in HB, the 
covariant approach is able to describe simultaneously the experimental data and LQCD results. Moreover, we have found that the consistency between both is improved from the good linear extrapolation obtained at LO (GMO) with the inclusion of the leading non-analytic terms. The success of a $SU(3)_F$-B$\chi$PT approach to describe current 2+1-flavor LQCD results may have important phenomenological applications, as has been shown with the determination of the $\sigma$ terms. An analysis of LQCD results obtained by other collaborations is in progress.

\section{Acknowledgments}

This work was partially supported by the  MEC contract  FIS2006-03438 and the EU Integrated Infrastructure Initiative Hadron Physics Project contract RII3-CT-2004-506078. JMC acknowledges the MICINN for support. LSG's work was partially supported by the Fundamental Research Funds for the Central Universities, the Alexander von Humboldt foundation, and the MICINN in the program Juan de la Cierva.


\begin{thebibliography}{99}

\bibitem{Jansen:2008vs}
  K.~Jansen,
  PoS {\bf LATTICE2008}, 010 (2008)
  [arXiv:0810.5634 [hep-lat]].

\bibitem{Gasser:1984gg}
  J.~Gasser and H.~Leutwyler,
  Nucl.\ Phys.\  B {\bf 250}, 465 (1985).

\bibitem{Gasser:1987rb}
  J.~Gasser, M.~E.~Sainio and A.~Svarc,
  Nucl.\ Phys.\  B {\bf 307}, 779 (1988).

\bibitem{Jenkins:1990jv}
  E.~E.~Jenkins and A.~V.~Manohar,
  Phys.\ Lett.\  B {\bf 255}, 558 (1991).

\bibitem{Fuchs:2003qc}
   J.~Gegelia and G.~Japaridze,
  Phys.\ Rev.\  D {\bf 60}, 114038 (1999); T.~Fuchs, J.~Gegelia, G.~Japaridze and S.~Scherer,
  Phys.\ Rev.\  D {\bf 68} (2003) 056005.

\bibitem{Pascalutsa:2000kd}
  V.~Pascalutsa,
  Phys.\ Lett.\  B {\bf 503}, 85 (2001).

\bibitem{Geng:2009hh}
  L.~S.~Geng, J.~Martin Camalich and M.~J.~Vicente Vacas,
  Phys.\ Lett.\  B {\bf 676}, 63 (2009).

\bibitem{Geng:2008mf}
  L.~S.~Geng, J.~Martin Camalich, L.~Alvarez-Ruso and M.~J.~V.~Vacas,
  Phys.\ Rev.\ Lett.\  {\bf 101}, 222002 (2008).


\bibitem{Geng:2009ys}
  L.~S.~Geng, J.~Martin Camalich and M.~J.~Vicente Vacas,
  Phys.\ Rev.\  D {\bf 80}, 034027 (2009).

\bibitem{Geng:2009ik}
  L.~S.~Geng, J.~Martin Camalich and M.~J.~Vicente Vacas,
  Phys.\ Rev.\  D {\bf 79}, 094022 (2009).

\bibitem{Geng:2010vw}
  L.~S.~Geng, N.~Kaiser, J.~Martin-Camalich and W.~Weise,
  Phys.\ Rev.\  D {\bf 82}, 054022 (2010).


\bibitem{MartinCamalich:2010fp}
  J.~Martin Camalich, L.~S.~Geng and M.~J.~Vicente Vacas,
  Phys.\ Rev.\  D {\bf 82}, 074504 (2010).


\bibitem{Jenkins:1991ts}
  E.~E.~Jenkins,
  Nucl.\ Phys.\  B {\bf 368}, 190 (1992).

\bibitem{Bernard:1993nj}
  V.~Bernard, N.~Kaiser and U.~G.~Meissner,
  Z.\ Phys.\  C {\bf 60}, 111 (1993).

\bibitem{Borasoy:1996bx}
  B.~Borasoy and U.~G.~Meissner,
  Annals Phys.\  {\bf 254}, 192 (1997).

\bibitem{WalkerLoud:2004hf}
  A.~Walker-Loud,
  Nucl.\ Phys.\  A {\bf 747}, 476 (2005).

\bibitem{Tiburzi:2004rh}
  B.~C.~Tiburzi and A.~Walker-Loud,
  Nucl.\ Phys.\  A {\bf 748}, 513 (2005).

\bibitem{Lehnhart:2004vi}
  B.~C.~Lehnhart, J.~Gegelia and S.~Scherer,
  J.\ Phys.\ G {\bf 31}, 89 (2005).

\bibitem{Pagels:1974se}
  H.~Pagels,
  Phys.\ Rept.\  {\bf 16}, 219 (1975).

\bibitem{Ellis:2008hf}
  J.~R.~Ellis, K.~A.~Olive and C.~Savage,
  Phys.\ Rev.\  D {\bf 77}, 065026 (2008).

\bibitem{Giedt:2009mr}
  J.~Giedt, A.~W.~Thomas and R.~D.~Young,
  Phys.\ Rev.\ Lett.\  {\bf 103}, 201802 (2009).

\bibitem{WalkerLoud:2008bp}
  A.~Walker-Loud {\it et al.},
  Phys.\ Rev.\  D {\bf 79}, 054502 (2009).

\bibitem{Aoki:2008sm}
  S.~Aoki {\it et al.}  [PACS-CS Collaboration],
  Phys.\ Rev.\  D {\bf 79}, 034503 (2009); $ibidem$ 
  Phys.\ Rev.\  D {\bf 81}, 074503 (2010).

\bibitem{Lin:2008pr}
  H.~W.~Lin {\it et al.}  [Hadron Spectrum Collaboration],
  Phys.\ Rev.\  D {\bf 79}, 034502 (2009).

\bibitem{Durr:2008zz}
  S.~Durr {\it et al.},
  Science {\bf 322}, 1224 (2008).

\bibitem{Bietenholz:2010jr}
  W.~Bietenholz {\it et al.},
  Phys.\ Lett.\  B {\bf 690}, 436 (2010).

\bibitem{Young:2009zb}
  R.~D.~Young and A.~W.~Thomas,
  Phys.\ Rev.\  D {\bf 81}, 014503 (2010).

\end{thebibliography}
\end{document}